\documentclass[conference]{IEEEtran}
\IEEEoverridecommandlockouts
\usepackage{cite}
\usepackage{amsmath,amssymb,amsfonts}
\usepackage{algorithmic}
\usepackage{graphicx}
\usepackage{tabularx}
\usepackage{multirow}
\usepackage{booktabs}
\usepackage{subcaption}
\usepackage[hyphens]{url}
\usepackage[bookmarksnumbered,unicode,hidelinks]{hyperref}
\usepackage{textcomp}
\usepackage{xcolor}

\def\BibTeX{{\rm B\kern-.05em{\sc i\kern-.025em b}\kern-.08em
    T\kern-.1667em\lower.7ex\hbox{E}\kern-.125emX}}
\begin{document}

\title{Making History Readable\\
}

\author{\IEEEauthorblockN{ Bipasha Banerjee*}
\IEEEauthorblockA{\textit{University Libraries} \\
\textit{Virginia Tech}\\
Blacksburg, VA USA  \\
0000-0003-4472-1902}
\and
\IEEEauthorblockN{Jennifer Goyne*}
\IEEEauthorblockA{\textit{University Libraries} \\
\textit{Virginia Tech}\\
Blacksburg, VA, USA \\
0009-0006-0037-5322}
\and
\IEEEauthorblockN{William A. Ingram}
\IEEEauthorblockA{\textit{University Libraries} \\
\textit{Virginia Tech}\\
Blacksburg, VA, USA \\
0000-0002-8307-8844} 
}

\maketitle
\def\thefootnote{*}\footnotetext{The authors contributed equally to this work}
\begin{abstract}
The Virginia Tech University Libraries (VTUL) Digital Library Platform (DLP) hosts digital collections that offer our users access to a wide variety of documents of historical and cultural importance.
These collections are not only of academic importance but also provide our users with a glance at local historical events.
Our DLP contains collections comprising digital objects featuring complex layouts, faded imagery, and hard-to-read handwritten text, which makes providing online access to these materials challenging.
To address these issues, we integrate AI into our DLP workflow and convert the text in the digital objects into a machine-readable format.
To enhance the user experience with our historical collections, we use custom AI agents for handwriting recognition, text extraction, and large language models (LLMs) for summarization.
This poster highlights three collections focusing on handwritten letters, newspapers, and digitized topographic maps. 
We discuss the challenges with each collection and detail our approaches to address them. 
Our proposed methods aim to enhance the user experience by making the contents in these collections easier to search and navigate.

\end{abstract}


\section{Introduction}
The DLP~\cite{dlp_web} is a cloud-native solution designed to accommodate collections, some as large as 40 TB.
Our collections include handwritten letters from the Civil War era, newspapers, and digitized historical maps.
Archival materials feature difficult-to-read 
handwriting, faded or irregular text, and complex layouts, making it challenging for digital libraries to provide online access.
These factors hinder accurate text recognition, complicate indexing and metadata generation, and obstruct full-text search functionality, ultimately reducing discoverability.
The DLP is designed to address these challenges on a large scale while organizing and preserving our extensive collections of complex digital materials.

The DLP integrates AI for processing challenging material, including text extraction to convert scanned documents, maps, and other materials into machine-readable form, custom AI agents for handwriting recognition, and large language models (LLMs) for summarization. 
We implement custom AI agents to extract text from handwritten documents, maps, and newspapers, making these difficult-to-read materials readable, thus enhancing the user experience. 
LLM-based summarization service enhances historical material by providing simplified summaries that are adapted for modern times.  
We discuss three collections: Silas Stepp handwritten 1860s Civil War letters, Montgomery Museum historic newspapers, and Virginia Tech's collection of Digitized Topographic Maps. 
We discuss the challenges with each collection and our solution that helps address them. Table~\ref{tab:collection} displays detailed information, such as size and item count, for the three collections discussed in the poster.
Our DLP facilitates easier access to the collection's rich resources, ultimately supporting academic research and fostering informed decision-making across various fields.
\begin{table}[htbp]
\caption{Collection Description}
\begin{center}
\label{tab:collection}
\begin{tabular}{p{1.7cm}p{1cm}rc}
\toprule
\textbf{Collection} &\textbf{Item Count} & 
\textbf{Size} & 
\textbf{Description}\\
\midrule
Silas Stepp &  4,070 & 414 MB & 1864 Civil War Letters [\ref{sec:silas}] \\
Montgomery Museum  & 1.44 Million & 23 TB & Historical Newspapers  [\ref{sec:newspaper}] \\
Digitized Topo. Maps & 12,178 & 3.87 TB & Series of Paper Maps [\ref{sec:newman_maps}]\\

\bottomrule
\end{tabular}
\end{center}
\end{table}

\section{Background}
Optical Character Recognition (OCR)~\cite{OCR} is a technology that aids in \textbf{text extraction} by converting the text in scanned images and documents into machine-readable form.
OCR can often result in noise in the extracted text due to inferior quality input images and the inability to process diversity in fonts and layouts, making it unreliable~\cite{ocr_problems}. 
Pytesseract~\cite{pytessaract} is a Python wrapper for Google's tesseract~\cite{tesseract} OCR engine that enables users to extract text from scanned documents. 
While Tesseract is a high-quality open-source solution for extracting text, it often fails to do very well with scanned documents and requires careful pre-processing.
AWS Textract~\cite{textract} is Amazon's proprietary machine learning service designed to extract text from handwriting, layout elements, and data from scanned or born-digital PDF/TIFF files or images.
It is a paid service that can be used with the AWS SDK, AWS web interface, or the AWS API key. 
The \verb|detect_document_text| API categorizes extracted text into AWS BlockType tags, identifying the structure of the document by recognizing elements such as pages, lines, and words.
The API provides us with extracted text, bounding-box information, and confidence scores for the associated block elements. 
Textract's \verb|analyze_document| API analyzes documents to extract and detect various elements, such as content, layout, style, and semantic elements. 

\textbf{Language Models}, specifically large language models (LLMs), perform exceptionally well on natural language generation tasks, such as automatic summarization.
Language models are built using the transformer architecture~\cite{NIPS2017_3f5ee243} that uses an attention mechanism that allows retention of contextual relationships across a sequence of texts.
Although language models such as BERT~\cite{devlin-etal-2019-bert} can also summarize, they are bound by context length limitation, which can be challenging when summarizing longer documents. 
Generative AI models such as Llama~\cite{llama-3}, GPT~\cite{GPT}, and Phi-3\cite{abdin_phi-3_2024} have improved their support for longer context lengths of up to 128k.
We use Meta's \verb|Llama-3.1-8B-Instruct| model to generate a summary of handwritten letters. 

\section{Use Cases}
\label{sec:use_cases}
\subsection{Handwritten images}
\label{sec:silas}

\begin{figure}[ht]
\centering
\includegraphics[width=0.8\linewidth]{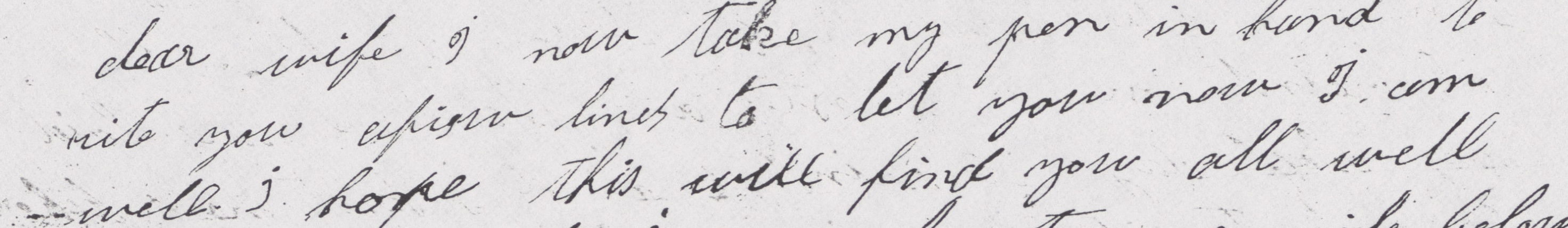}
\caption{An example from the Silas Stepp handwritten letters}
\label{fig:letter}
\end{figure}

\begin{figure}[ht]
\centering
\includegraphics[width=0.45\linewidth]{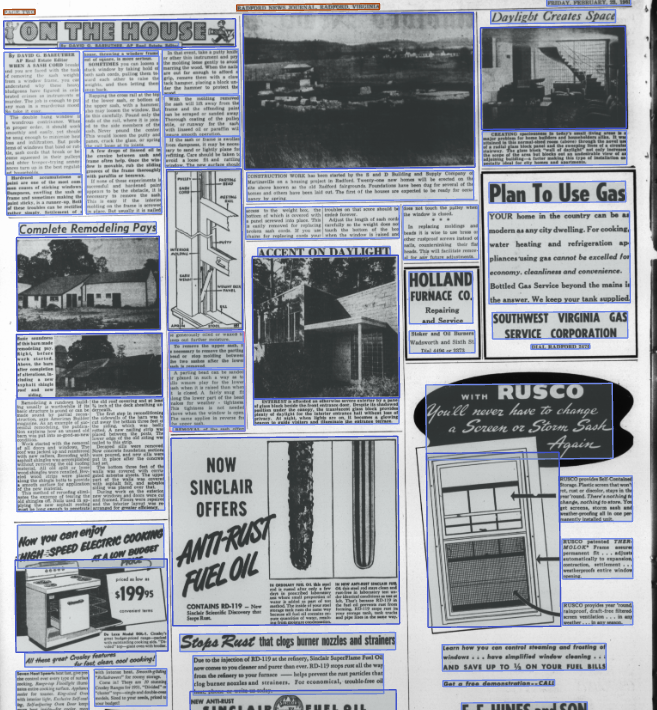}
\caption{Layout analysis on a newspaper image}
\label{fig:newspaper}
\end{figure}

\begin{figure}[ht]
\centering
\includegraphics[width=0.45\linewidth]{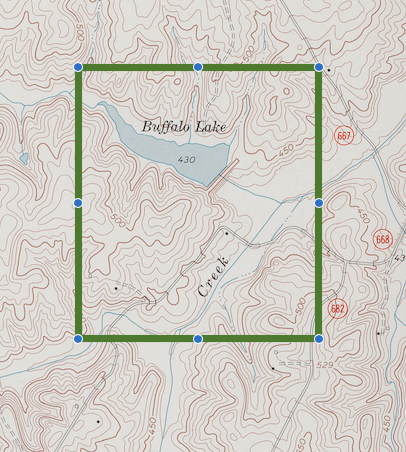}
\caption{A snapshot from the digitized maps collection}
\label{fig:maps}
\end{figure}

\begin{figure*}[ht]
\centering
\includegraphics[width=0.65\linewidth]{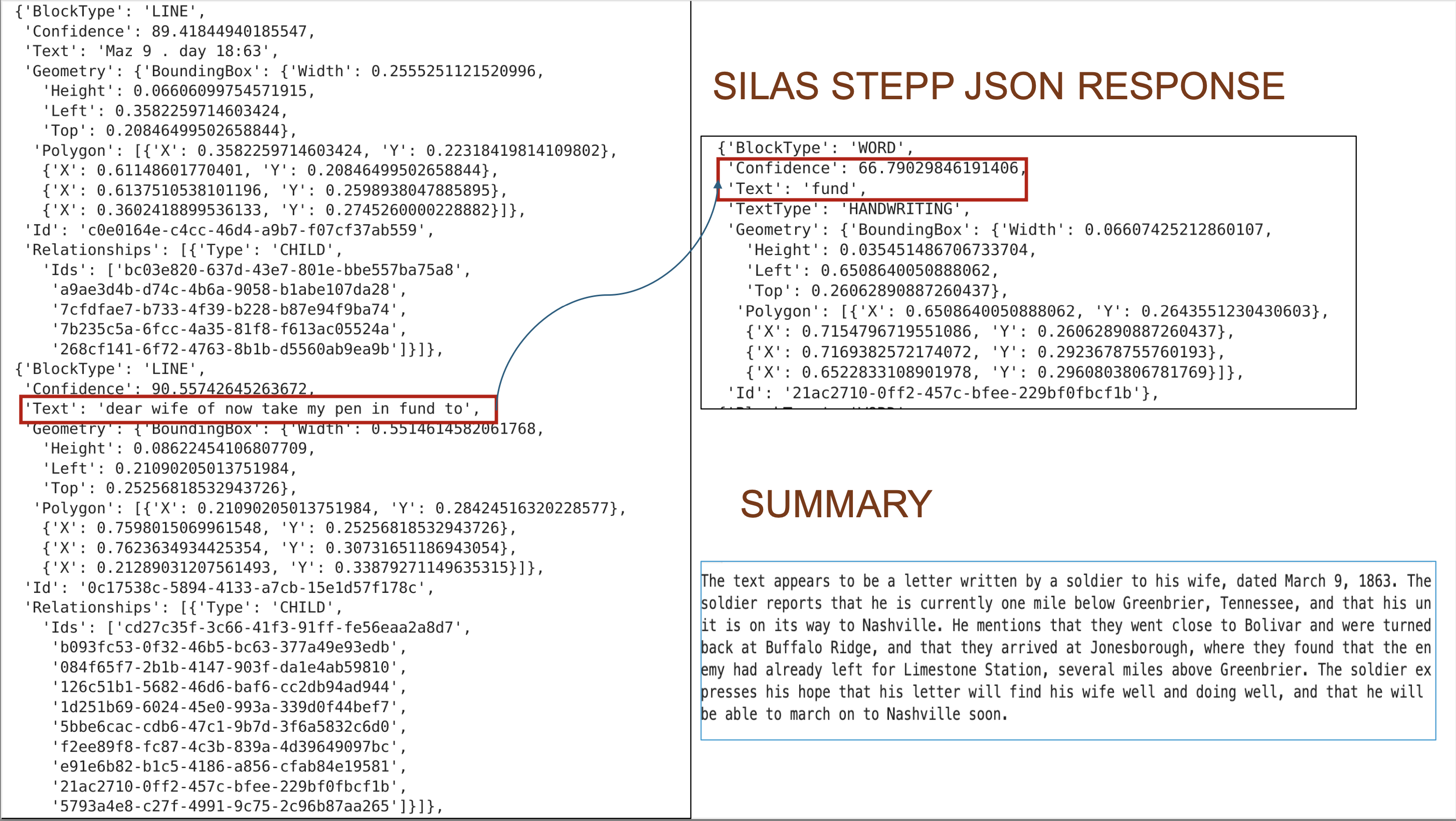}
\caption{Extracted text and corresponding summary from the Silas Stepp collection.}
\label{fig:extracted_text}
\end{figure*}

The letters feature cursive writing and archaic glyphs, which, combined with the overall quality of the digitized pages, make the text difficult to read (see Fig.~\ref{fig:letter}). 
Standard OCR relies on character-level segmentation, which is hard to do for handwritten pages due to the variability in letter shapes and contextual letter formulation that depends on surrounding words.

We added text extraction and summarization services to help make the content buried in these letters more accessible to the public. 
We addressed some inconsistencies in the extracted text as shown in Fig.~\ref{fig:extracted_text}.
The red box highlights an error in the extraction of the word `hand' as `fund' and the corresponding confidence score of 66.79\%.
To mitigate this challenge, we set a threshold for confidence scores that we deem acceptable.
If the confidence score for a block of text is below the set threshold, we utilize a language model to predict the top three most probable words and then select the best word. 
We use Google's \verb|bert-large-uncased-whole-word-masking| model for sentence correction. 
The model predicts the top three most probable words in a sentence where a particular word has a confidence score below the threshold. 
The text correction pipeline ensures that the most accurate text is used in summarization.  
Our summarization approach uses \verb|Llama-3.1-8B-Instruct| LLM to generate summaries from the historical letters. 
An example of the generated summary is shown in Fig.~\ref{fig:extracted_text}.
We simplify the words and summarize the content more attuned to current times, making it easier for our readers to understand historical handwritten letters.   

\subsection{Newspaper}
\label{sec:newspaper}

\textbf{Montgomery Museum}~\cite{newspaper} is a non-profit organization that houses a diverse array of historical content, such as physical artifacts, books, and newspapers detailing local and regional history in Virginia.
By offering this historical newspaper collection online in machine-readable format, users can search and navigate events from historical time periods.
The Montgomery Museum newspaper collection presents the text in a multi-column layout, where traditional OCR systems struggle with identifying the correct relative positioning of text in columns. 
The layout is further complicated by irregular column widths, overlapping text, and the presence of embedded images and advertisements. 
An example of applying layout analysis on a newspaper page is shown in Fig.~\ref{fig:newspaper}.

\subsection{Topographic Maps}
\label{sec:newman_maps}

\textbf{Virginia Tech Digitized Topographic Map Collection}~\cite{maps} is physically located at Virginia Tech University Libraries, which includes a series of maps that have undergone digitization for convenient digital access via the digital library platform. 
By making this collection online, users can view maps via the web, which previously could only be viewed in person with the assistance of a librarian to locate them physically.
These maps introduce challenges due to the non-linear placement of text, where the text is positioned at arbitrary angles or along curved paths around geographical features (see Fig.~\ref{fig:maps}). 
Traditional OCR models are optimized for rectilinear text, and text is expected to follow a predictable, left-to-right flow. When text is positioned at various angles or along curves, the image-preprocessing techniques required to correct orientation and alignment are not standard in traditional OCR systems, resulting in recognition errors. 
For this map collection, we first rotate the image at various angles to reorient the text and then apply Textract to each rotated version.
This multi-angle rotation strategy improves the overall success of text extraction.

\section{Discussion and Future Work}
This poster discusses custom AI agents for text extraction and summarization from historical digitized collections. 
By extracting text from challenging-to-read materials, we improve accessibility, while the summarization service creates clear, concise summaries to enhance user understanding, and thus improve overall engagement with the material.
These advances transform previously inaccessible content into searchable material and enrich the digital library platform. 

Our results discussed in Section~\ref{sec:use_cases} indicate that out-of-the-box solutions might need adjustments based on our DLP needs.
Our goal is to improve text extraction from the maps collection by using an ensemble method comprising tiling and rotation.
This approach involves processing large digital map objects and dividing them into several smaller tiles to accommodate and facilitate the file size requirements in the text extraction process.
We then apply rotation and text extraction to the smaller map tiles, ensuring consistent, high-quality results throughout the process.
We plan to integrate AI into services, such as automated metadata generation, that create richer and more descriptive metadata that improve categorization and organization. 
Automatic metadata generation speeds up the otherwise time-consuming process of metadata creation by using AI-driven approaches to identify and extract key elements from collections automatically. 
By strategically integrating AI into our DLP workflow, we are improving efficiency, enabling accurate large-scale extraction and summarization of complex materials and resulting in collections that are searchable, retrievable, and usable at scale.

\bibliographystyle{IEEEtran}
\bibliography{references.bib}
\end{document}